\documentstyle[12pt]{article}
   \newcommand{\nc}{\newcommand}
   \nc{\ds}{\displaystyle}
   \nc{\be}{\begin{equation}}
   \nc{\eq}{\end{equation}}
   \nc{\bc}{\begin{center}}
   \nc{\ec}{\end{center}}
   \nc{\bea}{\begin{eqnarray}}
   \nc{\eqa}{\end{eqnarray}}
   \nc{\non}{\nonumber}
   \nc{\tlift}{\rule[-2mm]{0mm}{6mm}}
   \nc{\slift}{\rule[-3mm]{0mm}{8mm}}
   \nc{\mlift}{\rule[-4mm]{0mm}{10mm}}
   \nc{\llift}{\rule[-5mm]{0mm}{12mm}}
   \nc{\al}{\alpha}
   \nc{\bt}{\beta}
   \nc{\gm}{\gamma}
   \nc{\dl}{\delta}
   \nc{\ep}{\epsilon}
   \nc{\vep}{\varepsilon}
   \nc{\zt}{\zeta}
   \nc{\kp}{\kappa}
   \nc{\lm}{\lambda}
   \nc{\sg}{\sigma}
   \nc{\vr}{\varphi}
   \nc{\om}{\omega}
   \nc{\Sg}{\Sigma}
   \nc{\Ps}{\Psi}
   \nc{\Ph}{\Phi}
   \nc{\Gm}{\Gamma}
   \nc{\vj}{\vec{\jmath}}
   \nc{\vp}{\vec{p}\hspace{0.2ex}}
   \nc{\vs}{\vec{s}\hspace{0.2ex}}
   \nc{\vM}{\vec{M}\hspace{0.2ex}}
   \nc{\cph}{\phi}
   \nc{\lph}{\Phi}
   \nc{\mph}{\varphi}
   \nc{\lra}{\longrightarrow}
   \nc{\lla}{\longleftarrow}
   \nc{\tl}{\tilde}
   \nc{\p}{\partial}
   \nc{\phat}{\widehat{p}}
   \nc{\qhat}{\widehat{q}}
   \nc{\pqhat}{\widehat{(p-q)}}
   \nc{\phati}{\widehat{2p}}
   \nc{\pll}{\parallel}
   \nc{\Dsl}{D\!\!\!\!/}
   \nc{\dsl}{d\!\!\!\!/}
   \nc{\hmu}{\hat{\mu}}
   \nc{\Real}{\mbox{\rm Re }}
   \nc{\RR}{ \mbox{\rule{0,10ex}{1,53ex}\hspace{-0.13ex}R} }
   \nc{\ZZ}{\Bbb{Z}}
   \nc{\NN}{\Bbb{N}}
   \nc{\lb}[1]{\label{#1} {\raggedleft{[\mbox{\tiny{#1}}]}}}

   \def\listoftables{\section*{List of Tables\@mkboth
          {LIST OF TABLES}{LIST OF TABLES}}\@starttoc{lot}}
   \def\listoffigures{\section*{List of Figures\@mkboth
          {LIST OF FIGURES}{LIST OF FIGURES}}\@starttoc{lof}}

   \catcode`\@=11
      \def\fnum@figure{\small {\bf Figure \thefigure}}
      \def\fnum@table{\small {\bf Table \thetable}}
   \catcode`\@=12
   \makeatletter
   \@addtoreset{equation}{section}
   \makeatother
   
\begin{document}
   \normalsize
   \noindent
\begin{flushright}January 1995\\PITHA 1/95\end{flushright}
\section*{   \begin{center}
Solutions of the Hamilton--Jacobi equation \\for one component\\ two
dimensional Field Theories
\end{center}}
\begin{center}
      {\large Wulf B\"ottger\footnote{e--mail:
       wulf@thphys.physik.rwth--aachen.de},
       Henning Wissowski\footnote{e--mail:
henning@acds16.physik.rwth--aachen.de
       }, Hans A.\ Kastrup\footnote{e--mail:
       kastrup@thphys.physik.rwth--aachen.de} \\
 Institut f\"ur Theoretische Physik E, RWTH Aachen, \\
 D--52056 Aachen, Fed.\ Rep.\ Germany}
\end{center}
\noindent
\begin{abstract}\footnotesize\noindent\bf
The Hamilton--Jacobi formalism generalized to 2--dimensional field theories
according to Lepage's canonical framework is applied to several covariant
real scalar fields, e.g.\ massless and massive Klein--Gordon,
Sine--Gordon, Liouville and $\phi^4$ theories. For simplicity we use
the Hamilton--Jacobi equation of DeDonder and Weyl.\\
Unlike mechanics we have to impose certain integrability conditions on the
velocity fields to guarantee the transversality relations between
Hamilton--Jacobi wave fronts and the corresponding families of extremals
embedded therein.\\
B\"acklund Transformations play a crucial role in solving the resulting
system of coupled nonlinear PDEs.
\vspace{2cm}\\
{\em To appear in: Proceedings of the XXVIIIth International Symposium
Ahrenshoop
on the Theory of Elementary Particles, Aug.\ 30 -- Sept.\ 3, 1994,
DESY--\-IfH (Zeuthen 1995)}
\end{abstract}
\normalsize
\newpage
\section{Introduction}
Varying the relativistic invariant action
integral leads to the still covariant Euler--Lagran\-ge formulation. However,
in getting a canonical Hamiltonian description, one has to break the manifest
covariance by distinguishing a time variable and regarding the other
Minkowski variables as ``indices'' counting an infinite number of degrees of
freedom. Treating field theories as one parametric ``mechanical'' Lagrangian
systems hides a part of their rich geometrical structure.
\\[2mm]
If one would like to avoid this by handling all the Minkowski variables on
the same footing following Cartan's framework of forms as applied by
Lepage\footnote{A reformulation of these ideas in the multisymplectic
framework may be found e.g.\ in \cite{go1}\cite{go2}.}
\cite{ka} it turns out that a variety of geometrically distinct Hamiltonian
formulations \cite{go2} exists, but only in theories which involve more than
one real field \cite{ka}\cite{ki}.
\\[2mm]
Nambu \cite{nambu1} and Hosotani \cite{hosotani1} postulated a quantum theory
for relativistic strings within this covariant formalism generalizing
semiclassical approximations . In
``conventional'' semiclassical considerations, where probability currents
associated with
families of extremals are considered, transversality relations between
solutions of the Hamilton--Jacobi and the canonical equations play a crucial
role.
But in contrast to mechanics the ability to embed extremals in a given
family of wave fronts can be maintained in field theories only if the
corresponding slope functions (velocity fields) satisfy certain
{\em integrability conditions} (IC).
\\[3mm]
The simplest examples for such a Hamilton--Jacobi theory
involves one dynamical field depending on two space--time (1+1)
variables. In this case it is not necessary to distinguish the different
canonical formulations mentioned above. We use the formalism
of DeDonder and Weyl for simplicity.
\\[2mm]
We consider the Ha\-mil\-ton--Ja\-co\-bi equation and the related
integrability condition for several 2--di\-men\-sio\-nal free and
selfinteracting models, e.g.\ massless and massive Klein--Gordon equations,
Sine--Gordon, Liouville and $\phi^4$ theory. For the corresponding system of
partial differential equations a family of solutions is constructed
perturbatively in such a way that a given single extremal of interest (e.g.\
a soliton like a (anti-) kink or a bell solution) can be embedded in sets of
wave fronts to be calculated from the Hamilton--Jacobi equation (HJE) and
the corresponding integrability condition. The expansion of the wave fronts
in powers of the field variable leads to a hierarchy of nonlinear PDEs that
can be reduced to linear PDEs with nonconstant coefficients. By applying
B\"acklund transformations they are reduced to wave or free Klein--Gordon
equations. Remarkably, after solving only two linear PDEs the general
solution for every order of the hierarchy is obtained by integration.
\\[2mm]
In this way one can construct a $d \!=\! 1$--parametric family of extremals
from one single solution of the canonical equations (strong embedding).

\section{Lepage's Canonical Formulation of Mechanics}
We very briefly recall Lepage's main idea of introducing the canonical
formalism in mechanics for one configuration variable $q$. The general case
is discussed in \cite{ka}.
\\[2mm]
The initial canonical Lagrangian form $\omega\!=\! L(t,q,\dot{q})\,\mbox{d}t$
is extended by the product of a Lagrangian multiplier $h(t,q,v)$ and the
Pfaffian form $\varrho\! = \!\mbox{d}q-v\,\mbox{d}t$ vanishing on the tangent
vectors of the extremals\footnote{As to the variational principle it is
preferable to consider the generalised velocity $v$ as an independent
variable. The vanishing of the differential form ${\varrho}$ ensures the
identification of $v(t)$ with $\dot{q}_0(t)$ on the extremals. $\varrho$
generates an ideal $I[\varrho]$ in the algebra $\Lambda$ of forms on the
extended configuration space ${\cal M}_{1+1}:=\{(t,q)\}$: $\forall\theta\in
\Lambda\,\forall\alpha\in I[\varrho]: \theta\wedge\alpha\in I[\varrho]$.}
${\cal C}\!=\!{\cal C}_0:=\left\{(t,q \!=\! q_0(t))\right\}$.
Thus the action integral ${\cal A}[{\cal C}]$ over the path ${\cal C}\!:=\!
\{(t,q(t))\}$ in the extended configuration space ${\cal M}_{1+1}:=
\{(t,q)\}$ is modified ${\cal A}[{\cal C}]\rightarrow\tilde{\cal A}[
{\cal C}]$ without changing the Euler--Lagrange equations and their
solutions ${\cal C}\!=\!{\cal C}_0$:
$$
{\cal A}[{\cal C}]=\int_{\cal C}\omega= \int_{\cal C} L\left(t,q(t),\dot{q}
(t)\right)\,\mbox{d}t
\quad\Longrightarrow\quad \tilde{\cal A}[{\cal C}]=\int_{\cal C}
\Omega=\int_{\cal C}\left[ L(t,q,v)\,\mbox{d}t+h(t,q,v)\,\varrho\right]\,.
$$
The Lagrangian multiplier $h(t,q,v)$ is fixed by the requirement that
a Hamilton--Jacobi theory exists. Hence $\Omega$ has to be locally exact:
$\Omega\!=\!\mbox{d} S(t,q)$ on every family of extremals covering ${\cal M
}_{1+1}\!=\!\{(t,q)\}$ (or a part of it). The resulting condition: $\mbox{d}
\Omega \in I[\varrho]$, namely
\begin{equation}
\mbox{d}\Omega=(\partial_v L-h)\,\mbox{d}v
\wedge{\mbox{d}}t+(\mbox{d}h-\partial_q L\,\mbox{d}t)\wedge\varrho=
\underbrace{(\partial_v L-h)}_{\textstyle =0!}{
\mbox{d}}v\wedge{\mbox{d}}t+0\left(\mbox{mod}I[\varrho]\right)\vspace{-2mm}
\end{equation}
leads to the standard definition of the canonical momentum: $p\! := \!h \!
\stackrel{!}{=} \!\partial_v L$, so that
\begin{equation}
\Omega=L\,\mbox{d}t+p\,\varrho=L\,\mbox{d}t+p\,(\mbox{d}q-v\,\mbox{d}t)=-(pv
-L)\mbox{d}t+p\,\mbox{d}q=-H\,\mbox{d}t+p\,\mbox{d}q\,.
\end{equation}
Thus the Legendre transformation $L \rightarrow H$, $v \rightarrow p$ can
be implemented as a change of the basis in the cotangent bundle ${\cal T}^{
\ast}({\cal M}_{1+1})$, $\varrho\rightarrow \mbox{d}q$, $\mbox{d}t
\rightarrow \mbox{d}t$.
\\[2mm]
The existence of a potential $S(t,q)$ for the basic differential form
$\Omega$ yields the familiar Hamilton--Jacobi equation for $S(t,q)$ and the
corresponding condition for the momentum:
\begin{equation}
\Omega=-H\left(t,q,p=\psi(t,q)\right){\mbox d}t+\psi(t,q)\,
\mbox{d}q\stackrel{\textstyle !}{=}\mbox{d}S(t,q)=
\partial_t S(t,q)\,\mbox{d}t+\partial_q S(t,q)\,\mbox{d}
q\,.
\end{equation}
Comparison of the coeffiecients yields:
\begin{equation}
\partial_t S(t,q)+H\left(t,q,p=\psi(t,q)\right)=0\,,\;\; p=\psi(t,q)=
\partial_q S(t,q)\,.
\end{equation}

\section{A Canonical Theory for Fields with one Component}
We consider the Lagrangian ${\cal L}$ as a function of one real scalar field
$\varphi$ (depending on the variables $z\!=\!(x\!+\!t)/2,\,\bar{z}\!=
\!(x\!-\!t)/2$) and the quantities $v, \bar{v}$ which on the extremals
coincide with the derivatives of the fields: $v\!=\!\partial_z \varphi_0,
\bar{v}\! =\! \partial_{\bar{z}}\varphi_0$.
\\[2mm]
As in mechanics the canonical 2--form $\omega\! = \!{\cal L}\,{ \mbox{d}}z
\wedge\mbox{d}\bar{z}$ is extended by two Lagrangian parameters $h(z,\bar{z},
\varphi)$, $\bar{h}(z,\bar{z},\varphi)$ and a 1--form $\varrho\!=\! \mbox{d}
\varphi-v\,\mbox{d}z-\bar{v}\,\mbox{d}\bar{z}$ that vanishes on the
2--dimensional extremals $\varphi\!=\! \varphi_0(z,\bar{z})$:
\begin{equation}
\Omega={\cal L}\,\mbox{d}z\wedge\mbox{d}\bar{z}+\bar{h}\,\mbox{d}z\wedge
\varrho+h\,\varrho\wedge\mbox{d}\bar{z}\,.
\end{equation}
The condition $\mbox{d}\Omega \in I[\varrho]$  --- resulting from the
requirement that a Hamilton--Jacobi theory exists ---  is:
$$
\mbox{d}\Omega=
(\partial_v{\cal L}-h)\mbox{d}v\wedge\mbox{d}z\wedge\mbox{d}\bar{z}+
(\partial_{\bar{v}}{\cal L}-\bar{h})\mbox{d}\bar{v}\wedge\mbox{d}z\wedge
\mbox{d}\bar{z}+0\left(\mbox{mod}I[\varrho]\right)\stackrel{!}{=}0\left(
\mbox{mod}I[\varrho]\right)\,.
$$
This leads to a determination of $h$, $\bar{h}$: $h \! = \! \partial_v {\cal
L}$ and $\bar{h}\! = \! \partial_{\bar{v}} {\cal L}$. As before the Legendre
transformation ${\cal L} \rightarrow {\cal H}$, $v \rightarrow h$, $\bar{v}
\rightarrow \bar{h}$ can be implemented as a change of the basis in the
cotangent bundle\footnote{${\cal M}_{2+1}$ denotes the extended configuration
space of two dimensional field theory: ${\cal M}_{2+1}\! := \! \{(z,\bar{z},
\varphi)\}$.} ${\cal T}^{\ast}({\cal M}_{2+1})$, $\varrho\rightarrow \mbox{d}
\varphi$, $\mbox{d}z \rightarrow \mbox{d}z$, $\mbox{d}\bar{z} \rightarrow
\mbox{d}\bar{z}$:
\begin{equation}
\Omega =-{\cal H}\,\mbox{d}z\wedge\mbox{d}\bar{z}+\bar{h}\,\mbox{d}z\wedge
\mbox{d}\varphi+h\,\mbox{d}\varphi\wedge\mbox{d}\bar{z}\,,\quad p:=h\,,\,
\bar{p}:=\bar{h}\; \mbox{and}\; {\cal H}=pv+\bar{p}\bar{v}-{\cal L}
\,.\label{omega}
\end{equation}
The choice of the Hamilton--Jacobi potentials $S$, $\bar{S}$ for making
$\Omega$ exact is no longer unique. In the case of DeDonder and Weyl it is
the following \cite{ri}:
\begin{equation}
\Omega=\mbox{d}\!\left\{S(z,\bar{z},\varphi)\,\mbox{d}
\bar{z}-\bar{S}(z,\bar{z},\varphi)\,\mbox{d}z\right\}=\mbox{d}S\wedge
\mbox{d}\bar{z}-\mbox{d}\bar{S}\wedge\mbox{d}z\,.
\end{equation}
Comparing this expression with equation (\ref{omega}) we obtain the
Hamilton--Jacobi equation and the conditions for the momenta for one
component fields in two dimensions:
\begin{equation}
\fbox{$\displaystyle \partial_z S+\partial_{\bar{z}} \bar{S}=-{\cal H}\,,
\quad\partial_{\varphi}S=p\, ,\quad \partial_{\varphi}\bar{S}=\bar{p}\, . $}
\vspace{2mm}
\end{equation}
In mechanics it is possible to construct wave fronts for $1$--parametric
families of extremals $q_0(t)$ that cover a certain region of the
configuration space \cite{ka} {\em and} vice versa. Given a solution of the
Hamilton--Jacobi equation (HJE), the so--called ``slope function"
\begin{equation}
\Phi(t,q) = \partial_p H(t,q,p\! = \!\partial_q S(t,q))
\end{equation}
determine the corresponding extremals $q_0(t)$ from the differential equation
\, $\dot{q}(t)\! = \!\Phi(t,q(t))$ \, the $1$--parametric solution of which
exists at least locally.
\\[2mm]
In general this is {\em not} true for field theories; the ability
to embed extremals $\varphi_0(z,\bar{z})$ in a given wave front can be
maintained only if the slope functions $v\!=\!\Phi(\varphi,z,\bar{z})$,
$\bar{v}\!=\!\bar{\Phi}(\varphi,z,\bar{z})$ obtained from the inverse
Legendre transformation
\begin{equation}
v=v(p=\partial_{\varphi}S,\bar{p}=\partial_{\varphi}\bar{S},z,\bar{z},\varphi)
\,,\quad \bar{v}=\bar{v}(p=\partial_{\varphi}S,\bar{p}=\partial_{\varphi}
\bar{S},z,\bar{z},\varphi)
\end{equation}
satisfy the integrability condition
\begin{equation}
\fbox{$\displaystyle
\frac{\mbox{d}}{\mbox{d}{\bar{z}}}\Phi(z,{\bar{z}},
\varphi(z,{\bar{z}})):=\partial_{\bar{z}}\Phi+\bar{\Phi}\cdot\partial_{\varphi}
\Phi\,=\,\frac{\mbox{d}}{\mbox{d}z}\bar{\Phi}(z,{\bar{z}},\varphi(z,{\bar{z}}
)):=\partial_z\bar{\Phi}+ \Phi\cdot\partial_{\varphi}\bar{\Phi}
\,.$}\label{g2}
\end{equation}

\section{Hamilton--Jacobi theory for one real field}
We here restrict ourselves to Lagrangian densities of the following type:
${\cal L}\!=\! \partial_z\varphi
\partial_{\bar{z}}\varphi\!-\!V(\varphi)$. The potential $V(\varphi)$ is an
analytic function. Here we have the canonical
momenta $p\!=\!\bar{v}$, $\bar{p}\!=\!v$, the Hamiltonian density ${\cal H}
\!=\!p\bar{p}+V$ and the slope functions $\Phi\!=\!\partial_{\varphi}
\bar{S}$, $\bar{\Phi}\!=\!\partial_{\varphi}S$. We have the
Hamilton--Jacobi equation
\begin{equation}
\partial_z S+\partial_{\bar{z}} \bar{S} = \partial_{\varphi} S\, \partial_{
\varphi} \bar{S}+V(\varphi)\label{HDW}
\end{equation}
and the related integrability condition
\begin{equation}
\partial_z\partial_{\varphi}S-\partial_{\bar{z}}\partial_{\varphi}\bar{S}
=\partial_{\varphi} S\, \partial_{\varphi}^2\bar{S}-
\partial_{\varphi} \bar{S}\, \partial_{\varphi}^2 S \,.\label{ICC}
\end{equation}
Knowing solutions $S$ and $\bar{S}$ of the equations (\ref{HDW}) and
(\ref{ICC})
a family of embedded extremals $\varphi=\tilde{\varphi}(z,\bar{z})$ is
determined by a system of first order PDEs:
\begin{equation}
\partial_z\tilde{\varphi}(z,\bar{z})=\tilde{\Phi}=\partial_{\varphi}\bar{S}(z,
\bar{z},\varphi=\tilde{\varphi}) \, ,\quad \partial_{\bar{z}}
\tilde{\varphi}(z,\bar{z})=\tilde{\bar{\Phi}}=\partial_{\varphi}S(z,\bar{z},
\varphi=\tilde{\varphi}).\label{qq}
\end{equation}
A solution is obtained by expanding $S(z,\bar{z},\varphi)$ and
$\bar{S}(z,\bar{z},\varphi)$ in powers of the difference $y\!=\! \varphi -
\varphi_0$ between $\varphi$ and a known extremal $\varphi_0(z,\bar{z})$:
\begin{equation}
S(z,\bar{z},\varphi)= \sum^{\infty}_{n=0} \frac{1}{n!} A_n(z,\bar{z}) y^n\,
, \quad \bar{S}(z,\bar{z},\varphi)= \sum^{\infty}_{n=0} \frac{1}{n!}
\bar{A}_n(z,\bar{z}) y^n\,.
\end{equation}
Naturally $\varphi_0$ has to satisfy the relation (\ref{qq}),
which fixes the functions\footnote{Obviously
$A_1$ and $\bar{A}_1$ satisfy the first order of the HJE and the zeroth
order of the IC.} $A_1\! = \! \partial_{\bar{z}} \varphi_0$ and $\bar{A}_1\!
= \! \partial_z \varphi_0 $ only, without influencing the remaining
coefficients. $A_0$ and $\bar{A}_0$ are only affected by the HJE of zeroth
order in $y$:
$
\partial_z A_0+\partial_{\bar{z}} \bar{A}_0 = \left.{\cal L}\right|_{\varphi=
\varphi_0} =  \partial_z\varphi_0
\partial_{\bar{z}}\varphi_0- 2V(\varphi_0).
$
Thus one of them can be chosen arbitrarily.
\\[2mm]
The IC of first order $\partial_z A_2 \! = \! \partial_{\bar{z}}\bar{A}_2$
permits to reduce these two functions at least locally to one generating
potential function: $A_2 \! = \! \partial_{\bar{z}} \ln(\theta) $ and
$\bar{A}_2 \! = \! \partial_z \ln(\theta) $. This logarithmic substitution
linearizes the HJE of order $y^2$:
\begin{equation}
\fbox{$\displaystyle\partial_{\bar{z}}\partial_z\theta(z,\bar{z})+\frac{1}{2}
\left\{\partial_{\varphi}^2V\left(\varphi\!=\!\varphi_0(z,\bar{z})\right)
\right\}\theta(z,\bar{z})=0\,.$}\label{variation}
\end{equation}
Though its coefficient is nonconstant the general solution of this PDE can be
obtained by employing B\"ack\-lund transformations.
\\[2mm]
The 1--parametric family of extremals in the vicinity ($y = \epsilon Y$,
$\epsilon\ll 1$) of the original extremal $\varphi_0(z,\bar{z})$ is
determined by integrating the PDEs (\ref{qq}):
\begin{equation}
\tilde{\varphi}(z,\bar{z},c_0)=\varphi_0(z,\bar{z})+ \epsilon c_0 \theta
(z,\bar{z})\,.
\end{equation}
The result is equivalent to that obtained by a second variation of the
action with respect to $\varphi$, where $\epsilon$
parametrizes a family of extremals.
\\[2mm]
In general the coefficients $A_n$, $\bar{A}_n$ are determined by the
$n$--th order of the HJE and the ($n\!-\!1$)th order of the IC\footnote{For
details see \cite{wi} and \cite{toll}.}:
$$
A_n= \theta^{1-n} \,\partial_{\bar{z}}
\left[\frac{\chi_n(z,\bar{z})}{\theta}-\bar{\chi}_n(z,\bar{z})\right]
\,,\quad
\bar{A}_n=\theta^{1-n}\, \partial_z
\left[\frac{\chi_n(z,\bar{z})}{\theta}+\bar{\chi}_n(z,\bar{z})\right]\, ,
\quad n\ge 3\,,
$$
where the infinite hierarchy of functions $\chi_n(z,\bar{z}),\bar{\chi}_n
(z,\bar{z})$ has to fulfil {\em only} two PDEs of second order:
\begin{eqnarray}
\partial_z\partial_{\bar{z}}\bar{\chi}_n(z,\bar{z})&=&
\mbox{Inhomogeneity}\,,\\
\partial_z\partial_{\bar{z}}\chi_n(z,\bar{z})+\frac{1}{2}\left(\partial_{
\varphi}^2V(\varphi_0)\right)\chi_n(z,\bar{z})&=&\mbox{Inhomogeneity}
\end{eqnarray}
with two inhomogeneities which depend on the functions $A_i,\bar{A}_i
$, $i\!=\!1,...n\!-\!1$.
\\[2mm]
{}From the general ansatz for the B\"acklund transformation \cite{miura}
\begin{equation}
\partial_z\hat{\theta}=F_1\left\{z,\bar{z},\theta,\hat{\theta},\partial_z
\theta\right\}\,,\qquad
\partial_{\bar{z}}\hat{\theta} = F_2\left\{z,\bar{z},
\theta,\hat{\theta},\partial_{\bar{z}}\theta\right\}
\end{equation}
and the requirement that $\hat{\theta}$ fulfills a free wave or Klein--Gordon
equation and the integrability condition $\partial_z
\partial_{\bar{z}}\hat{\theta}\!=\!\partial_{\bar{z}}\partial_z\hat{\theta}
\,\Rightarrow\,\mbox{d}F_1/\mbox{d}\bar{z}\!=\!\mbox{d}F_2/\mbox{d}z$,
we infer the relations
\begin{equation}
\partial_z\hat{\theta} =
+\partial_z\theta+\left(\partial_z\psi\right)(\hat{\theta}+
\theta)\,,\quad
\partial_{\bar{z}}\hat{\theta}=-\partial_{\bar{z}}\theta+\left(\partial_{
\bar{z}}\psi\right)(\hat{\theta}-\theta)\,.\label{back}
\end{equation}
Here $\psi\!=\!\psi(z,\bar{z})$ is required to be a {\em special} solution of
\begin{equation}
\partial_z\partial_{\bar{z}}\psi-(\partial_z\psi)(\partial_{\bar{z}}\psi)-
\frac{1}{2}
\partial_{\varphi}^2V(\varphi_0)=0\,,\quad \partial_z\partial_{\bar{z}}\psi+
(\partial_z\psi)(\partial_{\bar{z}}\psi)-m^2=0\,.
\end{equation}
Knowing $\psi$ and $\hat{\theta}$ we can calculate $\theta$ by integrating
the linear BTs (\ref{back}).

\section{Applications}
The Hamilton--Jacobi theory for a free scalar field
leads to the wave or the Klein--Gordon equation (\ref{variation})
without need of implementing a B\"acklund transformation (BT). Therefore we
address to the more interesting case of selfinteracting
fields. For more details of the following results see \cite{wi} and
\cite{toll}.

\begin{itemize}
\item
Applying our formalism to Liouville's theory ${\cal L}\!=\!\partial_z\varphi
\partial_{\bar{z}}\varphi+4\exp(\varphi)$ and using an arbitrary solution
of the corresponding equation of motion for which the general expression is
known \cite{miura},
\begin{equation}
\varphi_0\!=\!\ln\left\{2\frac{(\partial_z s(z))(\partial_{\bar{z}}
\bar{s}(\bar{z}))}{(s\!+\!\bar{s})^2}\right\}\,,
\end{equation}
the relation (\ref{variation}) yields
\begin{equation}
\partial_z\partial_{\bar{z}}\theta-2\frac{(\partial_z s)(\partial_{\bar{z}}
\bar{s})}{(s+\bar{s})^2}\theta=0\,,\quad\mbox{implying}\quad\partial_s
\partial_{\bar{s}}\hat{\theta}(s,\bar{s})=0
\end{equation}
by using the transformation $z\rightarrow s(z), \bar{z}
\rightarrow\bar{s}(\bar{z})$ and one BT $\psi\!=\!\ln(s\!+\!\bar{s})$.
\item
For the Sine--Gordon model ${\cal L}\!=\!\partial_z\varphi\partial_{\bar{z}}
\varphi\!+\!4[1\!-\!\cos(\varphi)]$ and starting from the (anti-) kink
solution $\varphi_0\!=\!\pm 4\arctan[\exp(z\!+\!\bar{z})]$ we
obtain:
\begin{equation}
\partial_z\partial_{\bar{z}}\theta-\{2\tanh^2(z+\bar{z})-1\}\theta=0\,.
\end{equation}
This can be reduced by one BT $\psi\!=\!\ln[\cosh(z\!+\!\bar{z})]$ to
a Klein--Gordon equation $\partial_z\partial_{\bar{z}}\hat{\theta}\!=\!
\hat{\theta}$. Then, using a Fourier transformation and integrating the
linear BT we obtain the general solution of (\ref{variation}).

\item
Contrary to the two models above the following ones can only be solved by at
least two BTs: the $\phi^4$--theories I) ${\cal L}\!=\!\partial_z
\varphi\partial_{\bar{z}}\varphi\!-\!4\varphi^2\!+\!2\varphi^4$ and
II) ${\cal L}\!=\!\partial_z\varphi\partial_{\bar{z}}\varphi\!+\!2\varphi^2\!-
\!2\varphi^4$. In both cases soliton
solutions are considered. We choose the (anti-) kink $\varphi_0\!=\!\pm
\tanh(z\!+\!\bar{z})$ and the bell solutions $\varphi_0\!=\!\pm\cosh^{-1}(z\!
+\!\bar{z})$, respectively. Then, we get the two relations
$$
\mbox{I)}\quad \partial_z\partial_{\bar{z}}\theta-\{6\tanh^2(z+\bar{z})-2\}
\theta=0,\quad\mbox{II)}\quad\partial_z\partial_{\bar{z}}\theta_2-\{6\tanh^2(
z+\bar{z})-5\}\theta_2=0\,.
$$
Both of them can be reduced to a free Klein--Gordon equation. The first
one by the two successive BTs $\psi_1\!=\!2\ln[\cosh(z\!+\!\bar{z})]$,
$\psi_2\!=\!\ln[\cosh(z\!+\!\bar{z})]\!+\!\dot{\imath}\sqrt{3}(\bar{z}\!
-\!z)$ and the second one by  using the BTs $\psi_3\!=\!2\ln[\cosh(z\!+\!
\bar{z})]\!+\!\sqrt{3}(\bar{z}\!-\!z)$ and $\psi_4\!=\!\ln[\cosh(z\!+\!
\bar{z})]$.
\end{itemize}

\end{document}